\documentclass[epj]{svjour}
\usepackage{graphics}
\begin{document}
\title{Fluctuations and Correlations in STAR}
\author{Brijesh K Srivastava, (for the STAR Collaboration)
}                     
\institute{Department of Physics\\
Purdue University \\	
West Lafayette, IN 47907, USA}
\date{Received: date / Revised version: date}
%
\abstract{
The study of correlations and fluctuations can provide evidence for the 
production of the quark-gluon plasma (QGP) in relativistic heavy ion 
collisions. Various theories predict that the production of a QGP phase 
in relativistic heavy ion collisions could produce significant 
event-by-event correlations and fluctuations in, transverse 
momentum, multiplicity, etc. Some of the recent results 
using STAR at RHIC will be presented along with results from other 
experiments at RHIC. The focus is on forward-backward multiplicity correlations, balance function, charge and transverse momentum fluctuations, and correlations.}

%
%
\maketitle
\section{Introduction}
\label{intro}
The investigation of high energy nucleus-nucleus collisions provides 
a unique tool to study the properties of hot and dense matter. The 
motivation is drawn from lattice QCD calculations, which predicts
a phase transition from hadro\-nic matter to a system of deconfined 
quarks and gluons  (QGP) at high temperature\cite{bib1}. The study of 
event-by- event fluctuations provides a novel probe to explore such 
transition in the search for the QGP. It is now  possible with 
large acceptance experiments at SPS and RHIC. 

\section{Forward-Backward Multiplicity Correlations}
\label{sec:1}

The measurement of particle correlations has been suggested as a method
to search for the existence of a phase transition in ultra-relativistic 
heavy ion collisions\cite{bib2,bib3,bib4}. If the quark-gluon plasma (QGP) 
is formed in these collisions, the existence or absence of particle 
correlations could lead to a determination of the presence of partonic 
degrees of freedom. A linear relationship has been found in high-energy 
colliding hadron experiments between the multiplicity in a forward $\eta$ 
region ($N_{f}$) and average multiplicity in a backward 
$\eta$ region ($N_{b}$)\cite{bib5,bib6}:

\begin{figure}
\centering        
\resizebox{0.45\textwidth}{!}{
\includegraphics{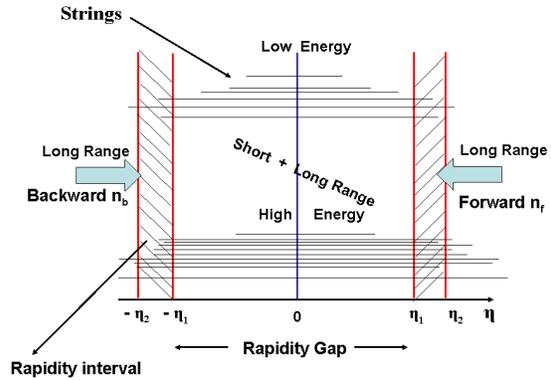}}
\caption{Pictorial representation of Forward-Backward correlations in pseudorapidity.}
\label{fig:1}       
\end{figure}

\begin{equation}\label{linear} 
\langle N_{b}(N_{f}) \rangle = a + bN_{f} 
\end{equation}
The coefficient b is referred to as the correlation coefficient and it can be expressed in terms of the expectation value \cite{bib5}:
\begin{equation}\label{b}
b = \frac{\langle N_{f}N_{b} \rangle- \langle N_{f} \rangle \langle N_{b} \rangle}{\langle N_{f}^{2} \rangle - \langle N_{f} \rangle^{2}} = \frac{D_{bf}^{2}}{D_{ff}^{2}}
\end{equation}
where $D_{bf}^{2}$ and $D_{ff}^{2}$ are the backward-forward and\\ forward-forward dispersions, respectively. This result is exact and model independent \cite{bib5}.
Short and long-range (in rapidity) multiplicity correlations are predicted as a signature of string fusion \cite{bib7,bib8}. The short range correlations are confined to midrapidity, while long range correlations are extended more than 2 units in rapidity. When strings fuse, a reduction in the long-range forward-backward correlation is expected. The existence of long-range multiplicity correlations may indicate the presence of multiple partonic inelastic collisions. These correlations arise from the superposition of a fluctuating number of strings, such that \cite{bib5}:
\begin{equation}\label{dbf}
\scriptsize{
\langle N_{f}N_{b} \rangle - \langle N_{f} \rangle \langle N_{b} \rangle \propto \left[\left(\langle n^{2} \rangle -\langle n \rangle ^{2}\right)\right]\langle N_{0f}\rangle \langle N_{0b}\rangle}
\end{equation}
with $(\left\langle n^{2}\right\rangle-\left\langle n \right\rangle^{2})$ the fluctuation in the number of inelastic collisions and $\left\langle N_{0f} \right\rangle, \left\langle N_{0b} \right\rangle$ the average multiplicity produced from a single inelastic collision.
Therefore, $D_{bf}^{2}$ should be sensitive to the presence of long-range multiplicity correlations.

We discuss the result on forward-backward multiplicity correlations from 200 GeV Au+Au collisions for all charged particles with $p_{T}$ in the range from 0.1 to 1.2 GeV/c, to ensure a sampling of soft particles only. To eliminate short-range correlations, a gap in pseudorapidity ($\eta$) of 1.6 units is considered \cite{bib4}. The forward pseudorapidity interval was $0.8 < \eta < 1.0$ and the backward was $-1.0 < \eta < -0.8$ \cite{tjt}. 

Fig. 2 shows the results for $D_{bf}^{2}$, $D_{ff}^{2}$, and the correlation coefficient, b, as a function of $N_{ch}$ for the 8 centrality bins. The presence of long-range multiplicity correlations are evident from $D_{bf}^{2}$ and b. The growth of $D_{bf}^{2}$ as a function of $N_{ch}$ is consistent with an increasing long-range correlation from peripheral to central heavy-ion collisions, corresponding to a greater number of fluctuating strings. 
\begin{figure}
\resizebox{0.45\textwidth}{!}{     
\includegraphics{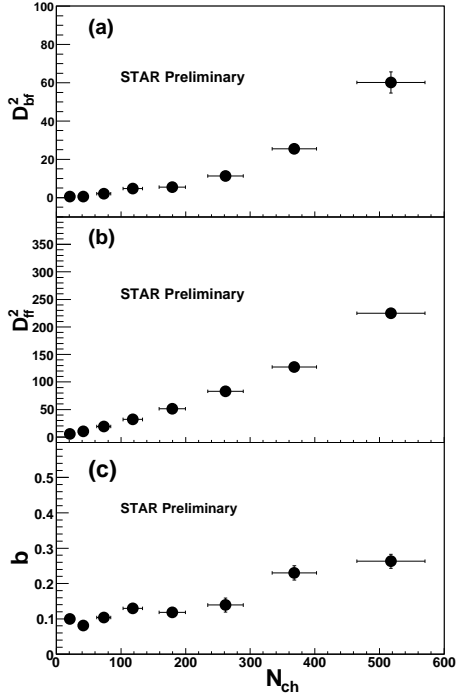}}
\caption{(a) $D_{bf}^{2}$, (b) $D_{ff}^{2}$, and (c) b from Au+Au collisions at $\sqrt{s_{NN}}$ = 200 GeV, as a function of the STAR reference multiplicity, $N_{ch}$.}
\label{fig:2}       
\end{figure}

The comparison of $D_{bf}^{2}$ and $D_{ff}^{2}$ as calculated from the Au+Au data to that from the Parton String Model (PSM) \cite{bib9,bib10} with the string fusion on or off is presented in Figs. 3 and 4, respectively. The PSM (with two string fusion) minbias multiplicity distribution closely matches that of the corrected Au+Au data. It also describes the $\left<p_{T}\right>$ enhancement, particle ratios, strangeness production, etc., seen in Au+Au collisions at RHIC \cite{bib9}. As such, the centrality cuts in the PSM correspond to the same percentage of the minimum bias cross section as in the data and are plotted as a function of the number of participant nucleons in the collision ($N_{part}$), calculated from Monte Carlo Glauber model \cite{bib11}. The statistical and systematic uncertainties in Figs. 3 and 4 are the same as those described for Fig. 2. 
\begin{figure}
\resizebox{0.45\textwidth}{!}{     
\includegraphics{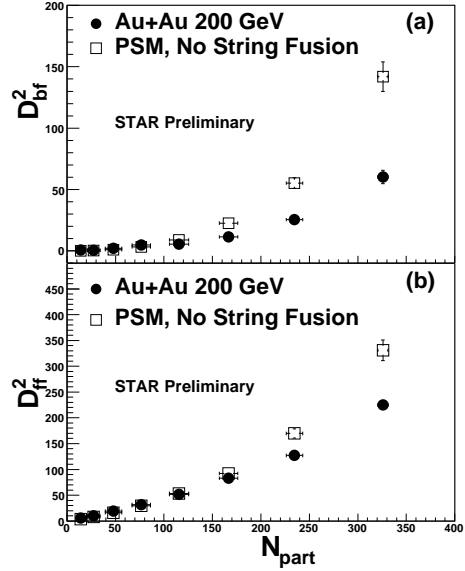}}
\caption{(a) $D_{bf}^{2}$ and (b) $D_{ff}^{2}$ as a function of the $N_{part}$ in 200 GeV Au+Au collisions, compared to the PSM with no string fusion (independent strings).}
\label{fig:3}       
\end{figure}
Figs. 3 and 4 show good agreement in peripheral collisions between data and the independent (no string fusion) or collective (with string fusion) model. In central Au+Au collisions, $D_{ff}^{2}$ with the independent string description deviates from the data, but shows good agreement for fusion of two soft strings. This confirms the agreement in multiplicity between the PSM (with two string fusion) and data. However, there is a large discrepancy in $D_{bf}^{2}$ for central collisions for both the independent and collective PSM, compared to the data. This discrepancy is greater for the case of independent strings (Fig. 3). This suggests an additional, dynamical reduction in the number of particle sources in central Au+Au collisions, greater than that provided by the fusion of two soft strings (Fig. 4). 

\begin{figure}
\resizebox{0.45\textwidth}{!}{      
\includegraphics{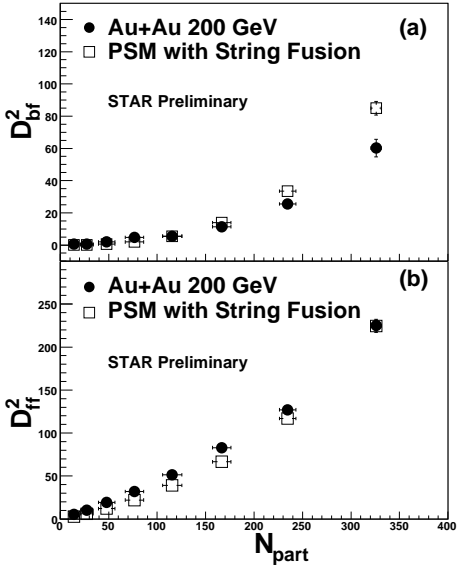}}
\caption{(a.) $D_{bf}^{2}$ and (b) $D_{ff}^{2}$ as a function of the $N_{part}$ in 200 GeV Au+Au collisions, compared to the PSM with string fusion (collective strings).}
\label{fig:4}       
\end{figure}
\section{Transverse Momentum Distributions and String Percolation}
\label{sec:2}
It is postulated that in the collision of two nuclei at high-energy, color strings are formed between projectile and target partons. These color strings decay into additional strings via  \begin{math} 
	q-\overline{q}
\end{math}
production, and ultimately hadronize to produce the observed hadron yields \cite{perc1}.
In the collision process, partons from different nucleons begin to overlap and form clusters in transverse space. The color strings are of radius r$_{0}$ = 0.20-0.25 fm \cite{perc1}. The fusion of strings to form clusters is an evolution of the Dual Parton Model \cite{bib6}, which utilizes independent strings as particle emitters, to the Parton String Model (PSM), which implements interactions (fusion) between strings \cite{perc2}. At some point, a cluster will form which spans the entire system. This is referred to as the maximal cluster and marks the onset of the percolation threshold. An overview of percolation theory can be found in the following reference \cite{perc3}. The quantity $\rho$, the percolation density parameter, can be used to describe overall cluster density. It can be expressed as 
\begin{equation}\rho = \frac{N\pi r_{0}^{2}}{S}\end{equation} 
with N the number of strings, S the total nuclear overlap area, and $\pi r_{0}^{2}$ the transverse disc area. At some critical value of $\rho = \rho_{c}$, the percolation threshold is reached. $\rho_{c}$ is referred to as the critical percolation density parameter. In two dimensions, for uniform string density, in the continuum limit, $\rho_{c}$ = 1.175 \cite{perc4}.
\\
To calculate the percolation parameter, $\rho$, a parameterization of pp events at 200 GeV is used to compute the $p_{T}$ distribution 
\begin{equation}\frac{dN}{dp_{T}^{2}} = \frac{a}{(p_{0}+p_{T})^{n}}\end{equation}
where a, $p_{0}$, and n are parameters fit to the data. This parameterization can be used for nucleus-nucleus collisions if one takes into account the percolation of strings by \cite{perc1}
\begin{equation}p_{0} \longrightarrow p_{0}\left( \frac{\left\langle \frac{nS_{1}}{S_{n}}\right\rangle _{Au-Au}}{\left\langle \frac{nS_{1}}{S_{n}}\right\rangle _{pp}}\right)^{\frac{1}{4}}\end{equation}
where $S_{1}$ and $S_{n}$ are the transverse overlap area produced by a single and N number of strings respectively.
In pp collisions at 200 GeV, the quantity $\left\langle \frac{nS_{1}}{S_{n}}\right\rangle _{pp} = 1.0 \pm 0.1$, due to low string overlap probability in pp collisions. Once the $p_{T}$ distribution for nucleus-nucleus collisions is determined, the multiplicity damping factor can be defined in the thermodynamic limit as \cite{perc5}
\begin{equation}\label{eq4}F(\rho) = \sqrt{\frac{1-e^{-\rho}}{\rho}}\end{equation}
which accounts for the overlapping of discs, with $1-e^{-\rho}$ corresponding to the fractional area covered by discs. 
\\
\begin{figure}
\centering
\resizebox{0.45\textwidth}{!}{      
\includegraphics{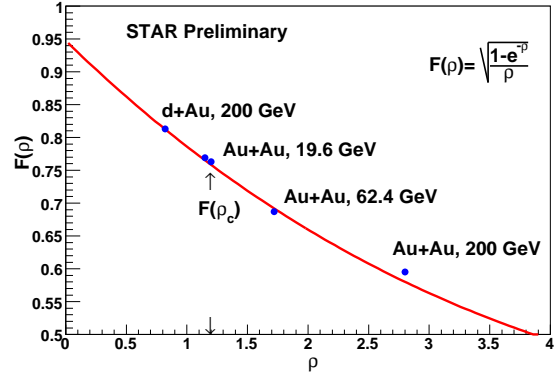}}
\caption{\footnotesize{Multiplicity suppression factor, F($\rho$) versus the percolation density parameter, $\rho$. The line is the function F($\rho$) and is drawn to guide the eye, not as a fit to the points. The estimated critical percolation density for 2-D overlapping discs in the continuum limit, $\rho_{c}$, is shown.}}
\label{fig:5}
\end{figure}
  
   The percolation density parameter, $\rho$, has been determined for several collision systems and energies. These results have been compared to the predicted value of the critical percolation density, $\rho_{c}$. If $\rho_{c}$ is exceeded it is expected that the percolation threshold has been reached, indicating the formation of a maximal cluster that spans the system under study. Figure 5 is a plot of the quantity F($\rho$) versus the percolation density parameter, ($\rho$), for central collisions. 
One can also consider the percolation density as a function of centrality in Au+Au collisions. The centrality expressed in terms of the number of participating nucleons ($N_{part}$) as found from Monte Carlo Glauber calculations \cite{tjt}. More central collisions correspond to greater values of $N_{part}$. Figure 6 shows $\rho$ as a function of the number of participant nucleons in Au+Au collisions at 200 and 62.4 GeV \cite{perc6}. For almost all collision centralities, 200 GeV Au+Au exceeds the critical percolation threshold, $\rho_{c}$. In 62.4 GeV Au+Au, all except the three most peripheral bins exceed $\rho_{c}$.
\begin{figure}
\centering
\resizebox{0.45\textwidth}{!}{      
\includegraphics{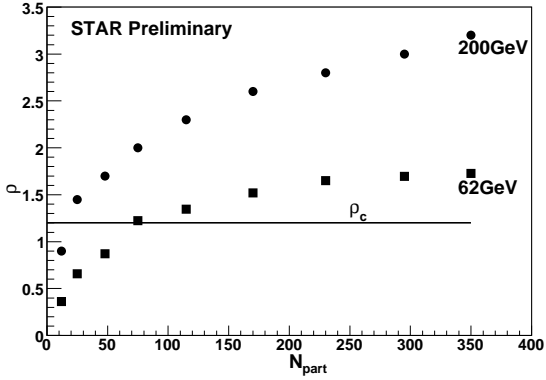}}
\caption{\footnotesize{The percolation density parameter, $\rho$, as a function of collision centrality ($N_{part}$) in 62.4 and 200 GeV Au+Au collisions. }}
\label{fig:6}
\end{figure}

\section{Balance Function}
\label{sec:3}
The balance function is based on the principle that charge is  locally conserved when particles are produced in pairs \cite{bal1,bal2}.
\begin{equation}
\scriptsize{
B(\Delta\eta) = \frac{1}{2} \left \{ \frac{N_{+-}(\Delta\eta)-N_{++}(\Delta\eta)}{N_{+}}+\frac{N_{-+}(\Delta\eta)-N_{--}(\Delta\eta)}{N_{-}}\right\}}
\end{equation}
where $N_{+-}$ is the number of charged pairs in a given pseudorapidity range, similarly for $N_{++}$, $N_{-+}$ and $N_{--}$. $N_{+}$ ($N_{-}$) is  the number of positive (negative) charged particles sum over all the events. $\Delta\eta=|\eta_{2}-\eta_{1}|$ is the width of the balance function.
If the system exists in a deconfined phase for an extended time and the charged pairs are produced at hadronization, the pairs  will retain more of their correlation in rapidity.
\begin{figure}
\centering
\resizebox{0.45\textwidth}{!}{      
\includegraphics{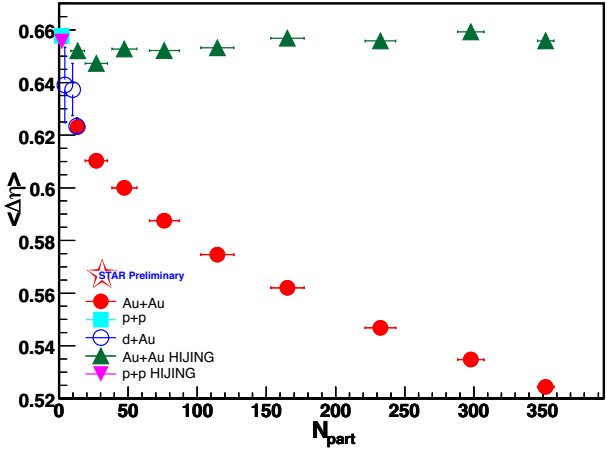}}
\caption{\footnotesize{The balance function widths for p+p, d+Au and Au+Au collisions at $\sqrt{(s_{NN})}=200$ GeV as a function of the number of participating nucleons along with HIJING calculations. }}
\label{fig:7}
\end{figure}
The balance function may be sensitive to whether the transition to a hadronic phase was delayed, as expected if the quark-gluon phase were to exist for a longer time.  In Fig. 7 the balance function widths for p+p, d+Au and Au+Au collisions at $\sqrt{(s_{NN})}=200$ GeV are shown as a function of the number of participants \cite{bal3}. The balance function widths scale with $N_{part}$.
The widths predicted using HIJING calculations are also shown in Fig. 7, along with data, and show 
little dependence on centrality and are similar to those measured for $\it pp$. The measured B$\Delta(\eta)$ in central Au+Au collisions is consistent with trends of model incorporating late hadronization. 

\section{Charge Fluctuations}
\label{sec:4}
Combined analysis of fluctuations in, e.g.,total and net charge 
for positively and negatively charged particles, as well as their ratios can reveal interesting physics. One of the quantity related to fluctuations is net charge fluctuations and is the difference of the number of produced positively and negatively charged particles measured in a fixed rapidity range, defined as \cite{char1,char2}
\begin{equation}
\nu_{+-} = \left \langle \left ( \frac{N_{+}}{\langle N_{+}\rangle} + \frac{N_{-}}{\langle N_{-}\rangle}\right )^2\right \rangle 
\end{equation}  
where $N_{+}$ and $N_{-}$ are multiplicities of positive and negative particles. The magnitude 
of the variance, $\nu_{+-}$, is determined both by statistical and dynamical fluctuations, $\nu_{+-}$ = $\nu_{+-,stat} + \nu_{+-,dyn}$. Details of net charge fluctuations analysis can be found in references \cite{char3}.
\begin{figure}
\centering
\resizebox{0.45\textwidth}{!}{      
\includegraphics{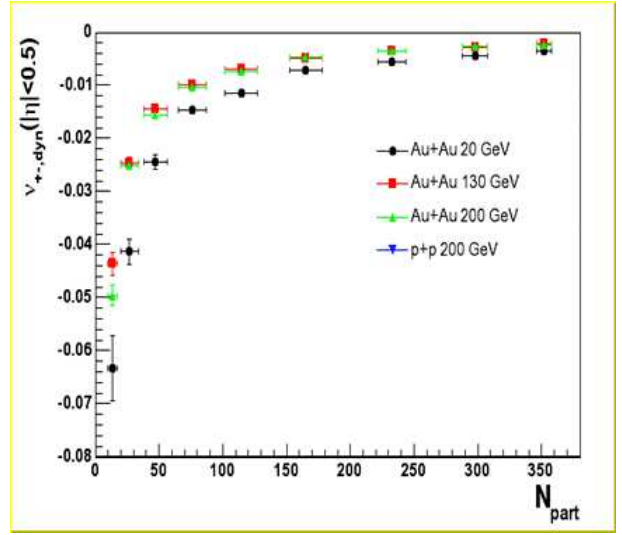}}
\caption{\footnotesize{$\nu_{+-,dyn}$ for all charged particles with $|\eta| <$ 0.5 from Au+Au collisions at 20, 130, and 200 GeV and pp at 200 GeV as a function of the number of participating nucleons.}}
\label{fig:8}
\end{figure}
  In Fig.8 the $\nu_{+-,dyn}$ is shown for Au+Au collisions at 20, 130, and 200 GeV along with the $\it pp$ data for all charged particles with $|\eta|\leq$0.5 and 0.1 $\leq p_{t} \leq$ 5 GeV/c as a function of number of participants \cite{char4}. One can see that $\nu_{+-,dyn}$ shows very little dependence on incident energy. 
\begin{figure}
\centering
\resizebox{0.45\textwidth}{!}{      
\includegraphics{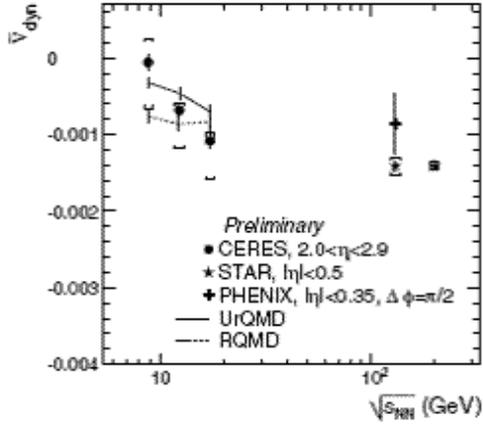}}
\caption{\footnotesize{$\nu_{+-,dyn}$ corrected for charge conservation as a function of collision energy}}
\label{fig:9}
\end{figure}
One must however correct measured $\nu_{+-,dyn}$ to account for charge conservation \cite{char4}. Fig. 9 show the plot of $\nu_{+-,dyn}$ = $\nu_{+-,dyn}$ + 4/$N_{ch}$ as a function of beam energy. The total charged particle multiplicity is given by $N_{ch}$.
Fig. 9 also shows the results from other experiments for comparison. 
The net charge fluctuations are smaller than expected based on predictions from a resonance gas or a quark gluon gas, which undergoes fast hadronization. 
\section{$p_{t}$ Fluctuations}
\label{sec:5}
Event-by-event fluctuations of mean $p_{t}$ have been proposed as a possible signature to search for the phase transition \cite{pt1,pt2,pt3}. Fluctuations involve a statistical component arising from the stochastic nature of particle production, as well as a dynamic component determined by correlations arising in various particle production processes. There are several measures to evaluate mean $p_{t}$ fluctuations. STAR has used 
$\left \langle \Delta p_{t,i}\Delta p_{t,j} \right \rangle$ \cite{pt4} and $\Delta\sigma_{pt}$ \cite{pt5} as the 
measure of the dynamical fluctuations. $\sum_{pT}$ gives the normalized fluctuation in CERES experiment \cite{pt6}. Another measure, $F_{pT}$, is defined as a deviation from 1 of the ratio of the rms of the event-by-event mean $p_{T}$ distribution in real events to that in mixed events(PHENIX) \cite{pt7}. 
To characterize transverse momentum correlation, the quantity  $\left \langle \Delta p_{t,i}\Delta p_{t,j} \right \rangle$ was calculated. Fig. 10 shows $\left \langle \Delta p_{t,i}\Delta p_{t,j} \right \rangle$ for Au+Au collisions at $\sqrt{s_{NN}}$ = 20, 62, 130 and 200 GeV. One observes that $\left \langle \Delta p_{t,i}\Delta p_{t,j} \right \rangle$ decreases with centrality. These results are also compared with HIJING calculations as shown in Fig. 10. The values for $\left \langle \Delta p_{t,i}\Delta p_{t,j} \right \rangle$ predicted by HIJING are always smaller than the data \cite{pt4}. 

\begin{figure}
\centering
\resizebox{0.45\textwidth}{!}{      
\includegraphics{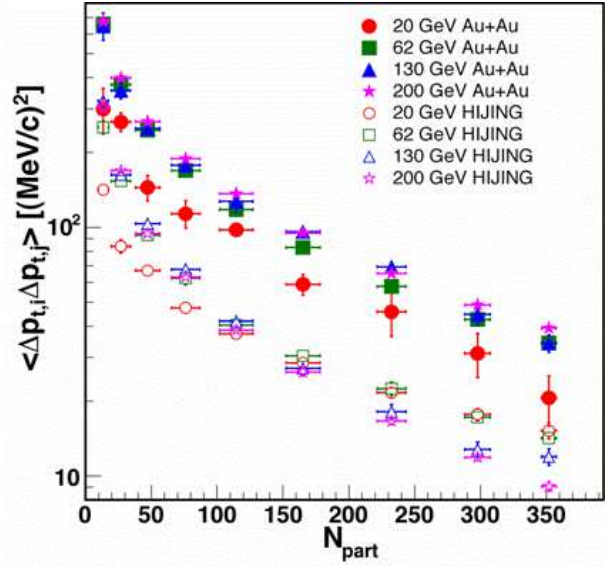}}
\caption{\footnotesize{$\langle \Delta p_{t,i} \Delta p_{t,j} \rangle$ as a function of centrality and incident energy for Au+Au collisions compared with HIJING results.}}
\label{fig:10}
\end{figure}
The pseudorapidity and azimuth ($\eta, \phi$) bin size dependence of event-wise mean transverse momentum fluctuations has been measured in terms of $\Delta\sigma_{pt}$ \cite{pt5,pt5x}. 
To access underlying dynamics, the corresponding autocorrelations ($\Delta\rho/\sqrt{\rho_{ref}}$) were extracted and are shown in Fig. 11 for 0-5\% central Au+Au collisions. The general form of the autocorrelations suggests that the basic correlation mechanism is parton fragmentation \cite{pt5x}.
\begin{figure}
\centering
\resizebox{0.35\textwidth}{!}{      
\includegraphics{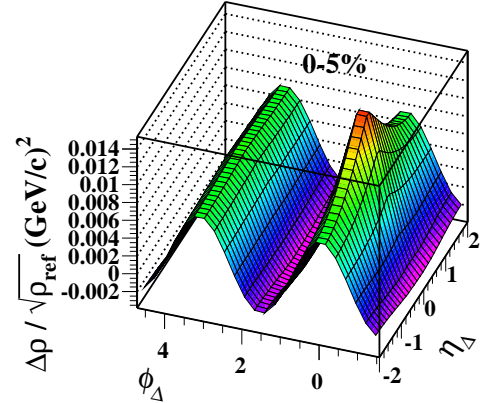}}
\caption{\footnotesize{ Autocorrelations, $\Delta\rho/\sqrt{\rho_{ref}}$ on difference variables ($\eta\Delta, \phi\Delta$) for 0-5\% central Au+Au collisions.}}
\label{fig:11}
\end{figure}


 $\sum_{pT}$ also measures the dynamical fluctuations and is proportional to mean covariance of all charged particle pairs per event. Fig. 12 shows $\sum_{pT}$ as a function of the nucleon-nucleon center of mass energy from SPS to RHIC energies. The observed fluctuations at SPS and at RHIC are about ~1\%. $F_{pT}$ is approximately proportional to $\langle N\rangle \sum_{pT}^{2}$, where $\langle N \rangle$ is the mean charged particle multiplicity. 
\begin{figure}
\centering
\resizebox{0.40\textwidth}{!}{      
\includegraphics{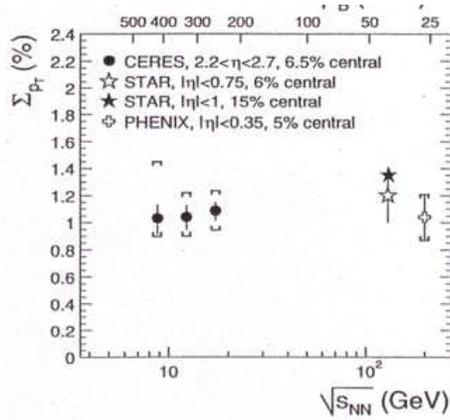}}
\caption{\footnotesize{$\sum_{pt}$ as a function of $\sqrt{s_{NN}}$ in central events.}}
\label{fig:12}
\end{figure}
Fig. 13 shows the magnitude of $F_{pt}$ as a function of centrality for Au+Au collisions with $p_{T}^{max}$ = 2.0 GeV/c. A significant non random fluctuation is seen that appears to peak in mid central collisions \cite{pt7}. 
\begin{figure}
\centering
\resizebox{0.40\textwidth}{!}{      
\includegraphics{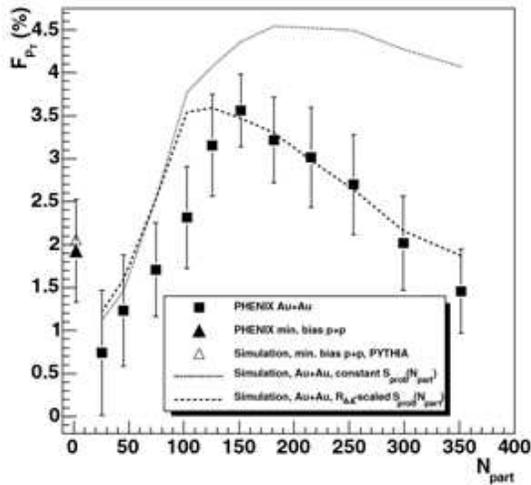}}
\caption{\footnotesize{$F_{pt}$ as a function of centrality. The result from PYTHIA
 simulation is also shown. }}
\label{fig:13}
\end{figure}
\section{Summary}
Some results from fluctuation and correlation measurements in STAR have been presented. 
The forward-backward multiplicity correlation in Au+Au shows collective behavior and a fusion/percolation approach has been explored\\ to understand this. 
The narrowing of balance function\\ width in central collisions is consistent
with trends \\predicted by models incorporating delayed hadronization.
The net charge dynamical fluctuations are found to be negative and
smaller than the value expected from a resonance or quark-gluon gas. 
Transverse momentum correlations and fluctuations show interesting physics which need further exploration.

\end{document}